\documentclass[journal]{IEEEtran}

\ifCLASSINFOpdf
  \usepackage[pdftex]{graphicx}
	\graphicspath{ {images/} }

\else

\fi

\usepackage{amsmath}

\usepackage{url}
\usepackage{cite}
\usepackage{blindtext}
\usepackage{caption}
\usepackage{subcaption}
\usepackage[dvipsnames]{xcolor}
\usepackage{amssymb}
\usepackage{enumitem}
\usepackage{lipsum}
\usepackage{mathtools}
\usepackage{mathtools, cuted}
\usepackage{lipsum, color}
\usepackage[margin=15mm]{geometry}
\usepackage{multicol}
\usepackage{float}
\usepackage{subcaption}

\floatstyle{plain}
\newfloat{twocolequfloat}{b}{zzz}
\floatname{twocolequfloat}{Equation}

\begin{document}

\title{Self-Interference Channel Characterization in Underwater Acoustic In-Band Full-Duplex Communications Using OFDM}

\author{Mohammad Towliat, Zheng Guo, Leonard J. Cimini, Xiang-Gen Xia and Aijun Song   % <-this % stops a space

\thanks{\noindent\rule{8.8cm}{0.4pt}

This work was supported primarily by the Engineering Research Centers Program of the National Science Foundation under NSF Cooperative Agreement No. (CNS-1704097, CNS-1704076). Any opinions, findings and conclusions or recommendations expressed in this material are those of the authors and do not necessarily reflect those of the National Science Foundation.

M. Towliat, L. J. Cimini and  X.-Gen Xia are  with the Department of Electrical and Computer Engineering, University of Delaware, Newark, DE, USA (e-mail: \{mtowliat, cimini, xianggen\}@udel.edu). A. Song, and Z. Guo are  with the Department of Electrical and Computer Engineering, University of Alabama, Tuscaloosa, AL, USA (e-mail: \{zguo18@crimson.ua.edu, song@eng.ua.edu\}).}% <-this % stops a space

}

% note the % following the last \IEEEmembership and also \thanks - 
% these prevent an unwanted space from occurring between the last author name
% and the end of the author line. i.e., if you had this:
% 
% \author{....lastname \thanks{...} \thanks{...} }
%                     ^------------^------------^----Do not want these spaces!
%
% a space would be appended to the last name and could cause every name on that
% line to be shifted left slightly. This is one of those "LaTeX things". For
% instance, "\textbf{A} \textbf{B}" will typeset as "A B" not "AB". To get
% "AB" then you have to do: "\textbf{A}\textbf{B}"
% \thanks is no different in this regard, so shield the last } of each \thanks
% that ends a line with a % and do not let a space in before the next \thanks.
% Spaces after \IEEEmembership other than the last one are OK (and needed) as
% you are supposed to have spaces between the names. For what it is worth,
% this is a minor point as most people would not even notice if the said evil
% space somehow managed to creep in.

% The paper headers
\markboth{}%
{Shell \MakeLowercase{\textit{et al.}}: Bare Demo of IEEEtran.cls for IEEE Journals}
% The only time the second header will appear is for the odd numbered pages
% after the title page when using the twoside option.
% 
% *** Note that you probably will NOT want to include the author's ***
% *** name in the headers of peer review papers.                   ***
% You can use \ifCLASSOPTIONpeerreview for conditional compilation here if
% you desire.

% If you want to put a publisher's ID mark on the page you can do it like
% this:
%\IEEEpubid{0000--0000/00\$00.00~\copyright~2015 IEEE}
% Remember, if you use this you must call \IEEEpubidadjcol in the second
% column for its text to clear the IEEEpubid mark.

% use for special paper notices
%\IEEEspecialpapernotice{(Invited Paper)}

% make the title area
\maketitle

% As a general rule, do not put math, special symbols or citations
% in the abstract or keywords.
\begin{abstract}
Due to the limited available bandwidth and dynamic channel, data rates are extremely limited in underwater acoustic (UWA) communications. Addressing this concern, in-band full-duplex (IBFD) has the potential to double the efficiency in a given bandwidth. In an IBFD scheme, transmission and reception  are performed simultaneously in the same frequency band. However, in UWA-IBFD,
 because of  reflections from the surface and bottom and the inhomogeneity of the water, a significant part of the transmitted signal returns back to the  IBFD receiver. This signal contaminates the desired signal from the remote end and is known as the self-interference (SI). With an estimate of the self-interference channel impulse response (SCIR), a receiver can estimate and eliminate the SI. A better understanding of the statistical characteristics of the SCIR is necessary for an accurate SI cancellation. In this article, we use an orthogonal frequency division multiplexing (OFDM) signal to characterize the SCIR in a lake water experiment. To verify the results, SCIR estimation is performed using estimators in both the frequency and time domains.
We show that, in our experiment, regardless of the depth of hydrophone, the direct path of SCIR is strong, stable and easily tracked; however, the reflection paths are weaker and rapidly time-varying making SI cancellation challenging. Among the reflections, the first bounce from the water surface is the prevalent path with a short coherence time around $70$ ms.
\end{abstract}

% Note that keywords are not normally used for peerreview papers.

% For peer review papers, you can put extra information on the cover
% page as needed:
% \ifCLASSOPTIONpeerreview
% \begin{center} \bfseries EDICS Category: 3-BBND \end{center}
% \fi
%
% For peerreview papers, this IEEEtran command inserts a page break and
% creates the second title. It will be ignored for other modes.
\IEEEpeerreviewmaketitle

\section{Introduction}
% The very fSCIRt letter is a 2 line initial drop letter followed
% by the rest of the fSCIRt word in caps.
% 
% form to use if the fSCIRt word consists of a single letter:
% \IEEEPARstart{A}{demo} file is ....
% 
% form to use if you need the single drop letter followed by
% normal text (unknown if ever used by the IEEE):
% \IEEEPARstart{A}{}demo file is ....
% 
% Some journals put the fSCIRt two words in caps:
% \IEEEPARstart{T}{his demo} file is ....
% 
% Here we have the typical use of a "T" for an initial drop letter
% and "HIS" in caps to complete the fSCIRt word.
\IEEEPARstart{U}{nderwater} acoustic (UWA) communication is an interesting research topic because of its wide applications such as monitoring the effect of human activities on marine life and forecasting disasters. However, due to limited available bandwidth \cite{R1}, UWA communication systems inherit a low data rate which cannot satisfy the rising demands in this field. Using in-band full-duplex (IBFD) transmission in UWA systems provides promise for improving the data rate, without increasing the bandwidth.
Generally, the main motivation for IBFD is to provide a higher spectrum efficiency by transmitting and receiving simultaneously in the same frequency band. Unfortunately, the transmitted signal in this scenario will corrupt the received signal, causing self-interference (SI) \cite{R2}.
The key challenge for IBFD schemes is the accurate cancellation of the SI. 

The feasibility of radio frequency-IBFD (RF-IBFD) has been confirmed by employing advanced SI cancellation methods \cite{R3, R4}.
Even though the idea of UWA-IBFD is similar to that of RF-IBFD, the SI cancellation task in UWA-IBFD is much more challenging. Consider a node in an UWA-IBFD network transmitting and receiving signals by using a transducer and a hydrophone, respectively. In addition to the direct path between the transducer and the hydrophone, the sea surface and sea bottom reflect a considerable portion of the transmitted signal toward the hydrophone and cause strong SI. In addition, the inhomogeneity and fluctuations of the water lead to  time-variations in the SI signal \cite{R24}.
Accordingly, eliminating the strong and time-varying SI is the bottleneck in UWA-IBFD, especially in shallow water which naturally has more reflections.

Despite the significant number of works on RF-IBFD, few works have studied UWA-IBFD. In \cite{R6}, a time-reversal transmission has been introduced to eliminate the SI. A multi-user UWA-IBFD using frequency division and code division multiple access (CDMA) are presented in \cite{R7} and \cite{R8}, respectively. In \cite{R5} a joint analog and digital adaptive SI cancellation is investigated.
The authors in \cite{R16} present a fully digital SI cancellation by using recursive least-squares (RLS) adaptive filters.
In \cite{R15} a two-stage method is proposed, in which the relatively stable part of the SI is eliminated  first, and the more rapidly time-varying residual part is separately treated at the second stage.  

From the previous works, it is clear that by better understanding the self-interference channel impulse response (SCIR), one can better cancel the SI \cite{R25}. In this regard, our main concern in this paper is to study the  statistical characteristics of the SCIR in UWA-IBFD. We employ  orthogonal frequency division multiplexing (OFDM) \cite{R11} to estimate the SCIR in a lake water. For verifying the achieved results, estimations are performed in both the frequency domain (FD) and the time  domain (TD). Using the estimated SCIR, statistical characteristics of the channel are obtained; including the power delay profile (PDP), the autocorrelation function (ACF) and the coherence time (COT). 
According to our measurements, regardless of the hydrophone's depth, the direct path is relatively stable and contributes to the principal part of the SI; on the other hand, the reflected paths are rapidly time-varying. Among all reflections, the first bounce from the water surface is the dominant path with a short COT. In the lake experiment, the COT of this path is determined to be around $70$ ms, which is even shorter than the OFDM block length. This means that tracking such a severe time-varying channel cannot be performed by using the UWA OFDM channel estimation methods proposed in \cite{R17, R18, R19}, which assume that the channel coefficients remain constant over at least one OFDM block. 

The rest of this paper is organized as follows. In Section II, we describe the OFDM system model and the methods we use to estimate PDP, ACF and COT from the experimental data. In Section III, the experimental steps and results are presented and discussed. Finally, Section VI summarizes the paper.

\section{SCIR estimation using OFDM}
In this section, we first present the OFDM system model and the signal structure that is used in the experiment. Then, we explain the FD and TD methods for estimating SCIR and calculating PDP, ACF and COT.

\vspace{-5pt}
\subsection{System Model}
\begin{figure*}[!t]
\centering
\includegraphics [width=7in]{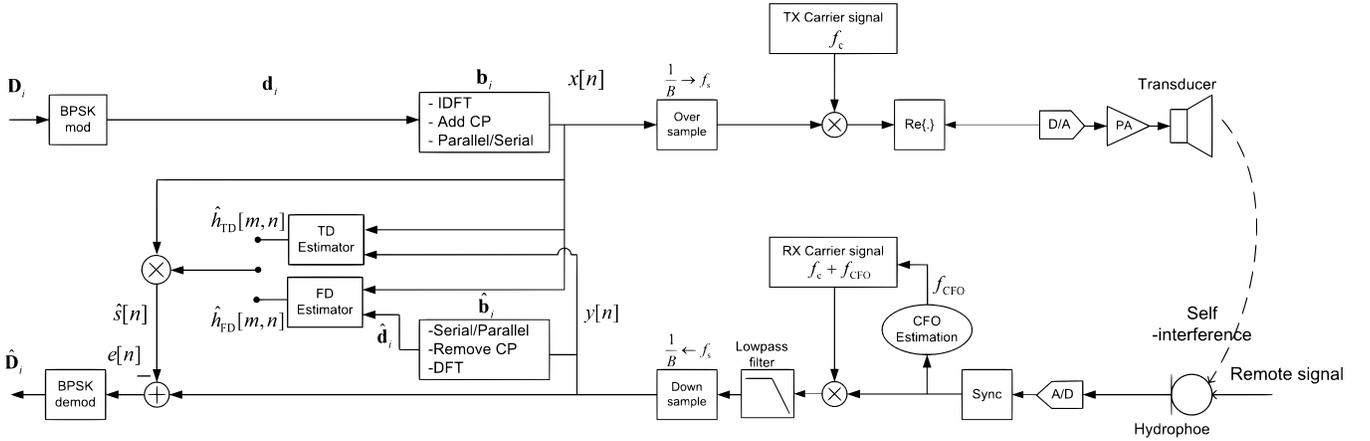}
\caption{UWA-IBFD system using OFDM for SCIR estimation, where $x[n]$ is the transmitted signal, $y[n]$ is the received signal, ${\hat s}[n]$ is the anticipated SI signal, and $e[n]$ denotes the residual. } \label{F1}
\end{figure*}

Due to the fact that the carrier frequencies and bandwidths are generally small in UWA communications, there exist analog-to-digital (A/D) converters that can sample the received signal much faster than the Nyquist  rate. Thus, in our formulations, we simply consider a discrete-time system model. Fig.~\ref{F1} illustrates the OFDM-IBFD system that we use in this work. 
As shown, $s[n]$ is the SI signal at time instance $n$. Assuming that the underwater environment behaves linearly, it is reasonable to model $s[n]$ as the convolution of the transmitted signal $x[n]$ and the SCIR, which could be time-varying in general \cite{R13}. Let us denote the time-varying SCIR by $h[m,n]$, where $m$ is the delay time and $n$ is the geotime. As a consequence, the SI signal can be represented  as \cite{R9}
\begin{equation}
s[n]=\sum\limits_{m=0}^{{M}-1}{h[m,n]x[n-m]}, \label{1}
\end{equation}

\noindent where ${M}$ is the SCIR length. In the absence of the remote end signal, the received signal is given as
\begin{equation}
y[n]=s[n]+w[n], \label{2}
\end{equation}

\noindent where $w[n]$ is the additive ambient noise. With an estimate of the SCIR, $\hat{h}[m,n]$, the receiver can estimate the SI as 
\vspace{-5pt}
\begin{align*}
\hat s[n] = \sum\limits_{m = 0}^{{M} - 1} {\hat h[m,n]x[n - m]},
\end{align*}

\noindent and the subtract it from the received signal, such that 
\begin{equation}
\begin{array}{*{20}{l}}
{e[n] = y[n] - \hat s[n]}\\
{{\mkern 1mu} {\mkern 1mu} {\mkern 1mu} {\mkern 1mu} {\mkern 1mu} {\mkern 1mu} {\mkern 1mu} {\mkern 1mu} {\mkern 1mu} {\mkern 1mu} {\mkern 1mu} \,\,\,\,\,\, =  \sum\limits_{m = 0}^{{M}-1} {(h[m,n] - \hat h[m,n]) x[n - m]}  + w[n]},
\end{array}
 \label{3}
\end{equation}

\noindent where ${e}[n]$ denotes the residual signal after interference cancellation. 
In UWA-IBFD, the ultimate goal is to cancel the SI down to the level of the ambient noise to make sure that its effect is eliminated.
According to \eqref{3}, the closer $\hat{h}[m,n]$ is to $h[m,n]$, the smaller the residual that remains. Since in UWA-IBFD the power of the SI is very high, sometimes even a relatively small  deviation between $\hat{h}[m,n]$ and $h[m,n]$ can result in a large residual, significantly degrading the performance of UWA-IBFD.
Based on this discussion, it is clear that the problem of accurate SI cancellation can be narrowed down to precise tracking of the SCIR, which in turn is based on a better understanding of the SCIR's statistical characteristics.

In order to statistically characterize the SCIR, the first step is to estimate the channel during a long period of geotime. In the experiment, we use an OFDM signal for channel estimation. 
There is a wide range of sophisticated UWA OFDM channel estimators \cite{R17, R18, R19}. Since the focus of this article is channel characterization, we simply employ the well-known FD and TD methods to estimate the SCIR; then, we use this estimate to extract the relevant statistical characteristics. 

\vspace{-5pt}
\subsection {OFDM Signal Structure}
According to Fig.~\ref{F1}, consecutive symbol vectors ${{\bf{D}}_i} = {[{D_i}[0], \ldots {D_i}[K - 1]]^T} \in {{\mathbb{C}}^{K \times 1}}$, for $i = 0, \ldots N-1$ are the inputs of the OFDM system in the frequency domain, where $K$ is the number of subchannels and $N$ is the number of transmitted OFDM blocks. The corresponding  vectors in the time domain are ${{\bf{d}}_i} = {[{d_i}[0], \ldots {d_i}[K - 1]]^T}\in {{\mathbb{C}}^{K \times 1}}$, which are the IDFTs of ${{\bf{D}}_i}$. For each $i$, a cyclic prefix (CP) with length $\upsilon$ is appended to ${{\bf{d}}_i}$ resulting in the OFDM block ${{\bf{b}}_i}= {[{b_i}[0], \ldots {b_i}[{L_{{\rm{blk}}}} - 1]]^T} \in {{\mathbb{C}}^{{L_{{\rm{blk}}}} \times 1}}$, where ${L_{{\rm{blk}}}}=K+\upsilon$ denotes the OFDM block length. We consider that the CP is the repetition of the end of the block appended to the head of the block, such that ${{\bf{b}}_i} = {[\left. {{d_i}[K - \upsilon], \ldots {d_i}[K - 1]} \,\, \right|{\bf{d}}_i^T]^T}$. For the purpose of studying the channel characteristics, we let $\upsilon  = K$; thus, the CP is an exact copy of the data part of the block.
Finally, the output of the OFDM transmitter, $x[n]$, is generated by serializing  the parallel vectors ${{\bf{b}}_i}$, such that 
\begin{equation} 
x[n] = {b_i}[l];\,\,\,n = i{L_{{\rm{blk}}}} + l,
 \label{5}
\end{equation}

\noindent  where $i = 0, \ldots N-1$ and $l = 0, \ldots {L_{{\rm{blk}}}}-1$.
Since  the main goal of this article is to provide the statistical characteristics of the SCIR, we simply assume that the entire OFDM signal is dedicated to channel estimation. 

\vspace{-5pt}
\subsection{Frequency Domain SCIR Estimation}

\noindent As shown in Fig.~\ref{F1}, for FD-SCIR estimation, the serial signal $y[n]$ is converted to parallel OFDM received blocks ${{{\bf{\hat b}}}_i} = {[{{\hat b}_i}[0], \ldots {{\hat b}_i}[{L_{{\rm{blk}}}} - 1]]^T}$, such that
\begin{equation}
{{\hat b}_i}[l] = y[n];\,\,\,n = i{L_{{\rm{blk}}}} + l.
\label{7}
\end{equation}

 \noindent Then, the CP is removed from ${{{\bf{\hat b}}}_i}$ giving ${{{\bf{\hat d}}}_i} = {[{{\hat d}_i}[0], \ldots {{\hat d}_i}[K - 1]]^T}$.  
Assuming that the channel remains constant (denoted by $h[m,i{L_{{\rm{blk}}}}]$) over the $i$th OFDM block, the DFT of ${{{\bf{\hat d}}}_i}$, ${{{\bf{\hat D}}}_i} = {[ {{{\hat D}_i}[0], \ldots {{\hat D}_i}[K - 1]} ]^T}$, is the element-wise multiplication of ${{{\bf{D}}}_i}$ by the channel frequency response \cite{R20}, so that
\begin{equation}
{{\hat D}_i}[k] = H[k,i{L_{{\rm{blk}}}}]\,{D_i}[k] + {W_i}[k],
\label{8}
\end{equation}

\noindent where
\vspace{-5pt}
\begin{align*}
H[k,i{L_{{\rm{blk}}}}] = \sum\limits_{m = 0}^{{M} - 1} {h[m,i{L_{{\rm{blk}}}}]{\mkern 1mu} \,{e^{ - j2\pi km/K}}}
\end{align*}

 \noindent is the $k$th DFT component of $h[m,i{L_{{\rm{blk}}}}]$ and ${W_i}[k]$ denotes the corresponding noise.
According to \eqref{8}, the estimation of $H[k,i{L_{{\rm{blk}}}}]$, for $k=0,\ldots K-1$, can be achieved by the element-wise division of  ${{\hat D}_i}[k]$ to ${D_i}[k]$ as
\begin{equation}
\hat H[k,i{L_{{\rm{blk}}}}] = \frac{{{{\hat D}_i}[k]}}{{{D_i}[k]}}. 
\label{9}
\end{equation}

\noindent Eq. \eqref{9} presents the estimated frequency components of the SCIR. Thus, the SCIR is obtained by taking the IDFT of $\hat H[k,i{L_{{\rm{blk}}}}]$,
\begin{equation}
{{\hat h}_{{\rm{FD}}}}[m,i{L_{{\rm{blk}}}}] = \sum\limits_{k = 0}^{K - 1} {\hat H[k,i{L_{{\rm{blk}}}}]\,{e^{j2\pi km/K}}} ,
\label{10}
\end{equation} 

\noindent where ${{\hat h}_{{\rm{FD}}}}[m,i{L_{{\rm{blk}}}}]$ denotes the estimation of the SCIR using the FD method, which is separately performed for each received OFDM block.
\vspace{-10pt}
\subsection{Time Domain SCIR Estimation}
In the previous section, we explained SCIR estimation in the frequency domain. One can also use TD methods on the OFDM signal, to verify the FD estimation results. 
As shown in Fig.~\ref{F1}, the received signal $y[n]$ is directly used for this purpose. We use the least squares (LS)\footnote{In order to reduce the noise effect, usually regularized LS estimations, such as the ridge method, are used for UWA channel estimation. However, since the noise is small compared to the SI signal, we consider the primary LS method.}
 method, in which the sliding observation window at time instance $n$ is defined as ${\bf{y}}[n] = {[y[n], \ldots y[n - {L_{{\rm{win}}}} + 1]]^T}\in {{\mathbb{C}}^{{L_{{\rm{win}}}} \times 1}}$, where ${L_{{\rm{win}}}}$ denotes the window size. Assuming that the channel remains constant over each observation window, according to \eqref{2}, the observed signal vector can be presented as
\begin{equation}
{\bf{y}}[n] = {{\bf{\bar X}}}[n]{\bf{h}}[n] + {\bf{w}}[n],
\label{11}
\end{equation}

\noindent where $\mathbf{h}[n]={{[h[0,n],\ldots h[{M}-1,n]]}^{T}}\in {{\mathbb{C}}^{{M} \times 1}}$ is the channel vector and ${{\bf{\bar X}}}[n]\in {{\mathbb{C}}^{{L_{{\rm{win}}}} \times {M}}}$ is the transmitted signal matrix defined as
\begin{equation}
\begin{array}{l}
{{\bf{\bar X}}}[n] = \\
\left[ {\begin{array}{*{20}{c}}
{x[n]}&{x[n - 1]}& \cdots &{x[n - {M} + 1]}\\
{x[n - 1]}&{x[n - 2]}& \cdots &{x[n - {M}]}\\
 \vdots &{}& \ddots & \vdots \\
{x[n - {L_{{\rm{win}}}} + 1]}&{}& \cdots &{x[n - {L_{{\rm{win}}}} - {M} + 2]}
\end{array}} \right].
\end{array}
\label{12}
\end{equation}

Assuming that ${L_{{\rm{win}}}} \ge  {M}$, the LS estimation of the SCIR in the time domain is
\begin{equation}
{{{{\bf{\hat h}}}_{{\rm{TD}}}}}[n] = {({{{\bf{\bar X}}}^H}[n]{{\bf{\bar X}}}[n])^{ - 1}}{{{\bf{\bar X}}}^H}[n]{\bf{y}}[n],
\label{13}
\end{equation}

\noindent where ${{{\bf{\hat h}}}_{{\rm{TD}}}}[n] = {[{{\hat h}_{{\rm{TD}}}}[0,n], \ldots {{\hat h}_{{\rm{TD}}}}[{M} - 1,n]]^T}$ is the estimated channel vector using  the TD method. To guarantee that, in \eqref{13}, the singularity of ${{{\bf{\bar X}}}^H}[n]{{\bf{\bar X}}}[n]$ is very unlikely, we take ${L_{{\rm{win}}}} =2{M}$. In contrast to the FD method, in the TD method, ${\bf{y}}[n]$ is a sliding observation window and \eqref{13} provides an estimation for each received sample.

\vspace{-5pt}
\subsection{Power Delay Profile Estimation}
One of the most relevant characters of a channel is the power delay profile (PDP) defined as ${{p[m]}} = {{\rm{E}}} \{ {{{\left| {{{h}}[m,n]} \right|}^2}} \} $, where ${{\rm{E}}}\left\{.\right\}$ denotes the expected value of the variable \cite{R13}. By transmitting $N$ OFDM blocks (each with length $L_{{\rm blk}}$), the FD method provides $N$ estimations of the SCIR; however, the TD method provides $\mathcal{N}=N{L_{{\rm{blk}}}}$ estimations by using the sliding observation window. 
According to the notations we have used in \eqref{10} and \eqref{13}, the estimated PDPs, by using the FD and TD methods, are respectively given as
\begin{equation}
\begin{array}{l}
{{{{\hat p}}}_{{\rm{FD}}}}[m] = \frac{1}{N}\sum\limits_{i = 0}^{N-1} {{{\left| {{{{{\hat h}}}_{{\rm{FD}}}}[m,i{L_{{\rm{blk}}}}]} \right|}^2}} ,\\
\\
{{{{\hat p}}}_{{\rm{TD}}}}[m] = \frac{1}{\mathcal{N}}\sum\limits_{n = 0}^{\mathcal{N} - 1} {{{\left| {{{{{\hat h}}}_{{\rm{TD}}}}[m,n]} \right|}^2}} .
\end{array}
\label{14}
\end{equation}

\vspace{-5pt}
\subsection{Autocorrelation Function and Coherence Time Estimation}

The other important statistical characteristics for the SCIR are the autocorrelation function (ACF) and the coherence time (COT). The ACF of the $m$th path of SCIR is defined as $q [m,{\eta}] = {{\rm{E}}}\left\{ {h[m,n]{\mkern 1mu} {h^*}[m,n - {\eta}]{\mkern 1mu} } \right\}$. Since $q [m,{\eta}]$ is generally a complex value with the maximum amplitude ${q[m,0]}$, usually the normalized ACF, $ {\bar q[m,\eta ]}  = \left| {q[m,\eta ]} \right|/q[m,0]$, is used to study the channel variation.
The COT, $\eta _{{\rm{CO}}}$, is also an interval for which $\bar q[m,\frac{{{\eta _{{\rm{CO}}}}}}{2}] > \mu $. In the literature, different values for $\mu$ is considered; however, we take $\mu=0.8$ \cite{R10}.

Given the estimated SCIRs in \eqref{10} and \eqref{13}, the estimations of the ACF are given as
\begin{equation}
\begin{array}{l}
{{{{\hat q}}}_{{\rm{FD}}}}[m,\eta {L_{{\rm{blk}}}}] = \frac{1}{N}\sum\limits_{i = 0}^{N-1} {{{{{\hat h}}}_{{\rm{FD}}}}[m,i{L_{{\rm{blk}}}}]\,{{\hat h}}_{{\rm{FD}}}^*[m,(i - \eta ){L_{{\rm{blk}}}}]} ,\\
\\
{{{{\hat q}}}_{{\rm{TD}}}}[m,\eta ] = \frac{1}{\mathcal{N}}\sum\limits_{n = 0}^{\mathcal{N} - 1} {{{{{\hat h}}}_{{\rm{TD}}}}[m,n]\,{{\hat h}}_{{\rm{TD}}}^*[m,n - \eta ]} .
\end{array}
\label{15}
\end{equation}

\noindent From \eqref{15}, it is obvious that, for the FD estimator, the ACF resolution is ${L_{{\rm{blk}}}}$ samples; however, using the TD method provides the ACF with an accuracy of one sample.

\section{Lake experiment}
In the previous section, we presented the relevant SCIR characterizations and the methods we use to estimate them in the experiment. In this section, we describe the lake experiment and the statistics we obtained.  

\vspace{-5pt}
\subsection{Experimental Set-up}
\begin{figure}
\centering
\begin{minipage}{.25\textwidth}
  \centering
  \includegraphics[height=2.3in]{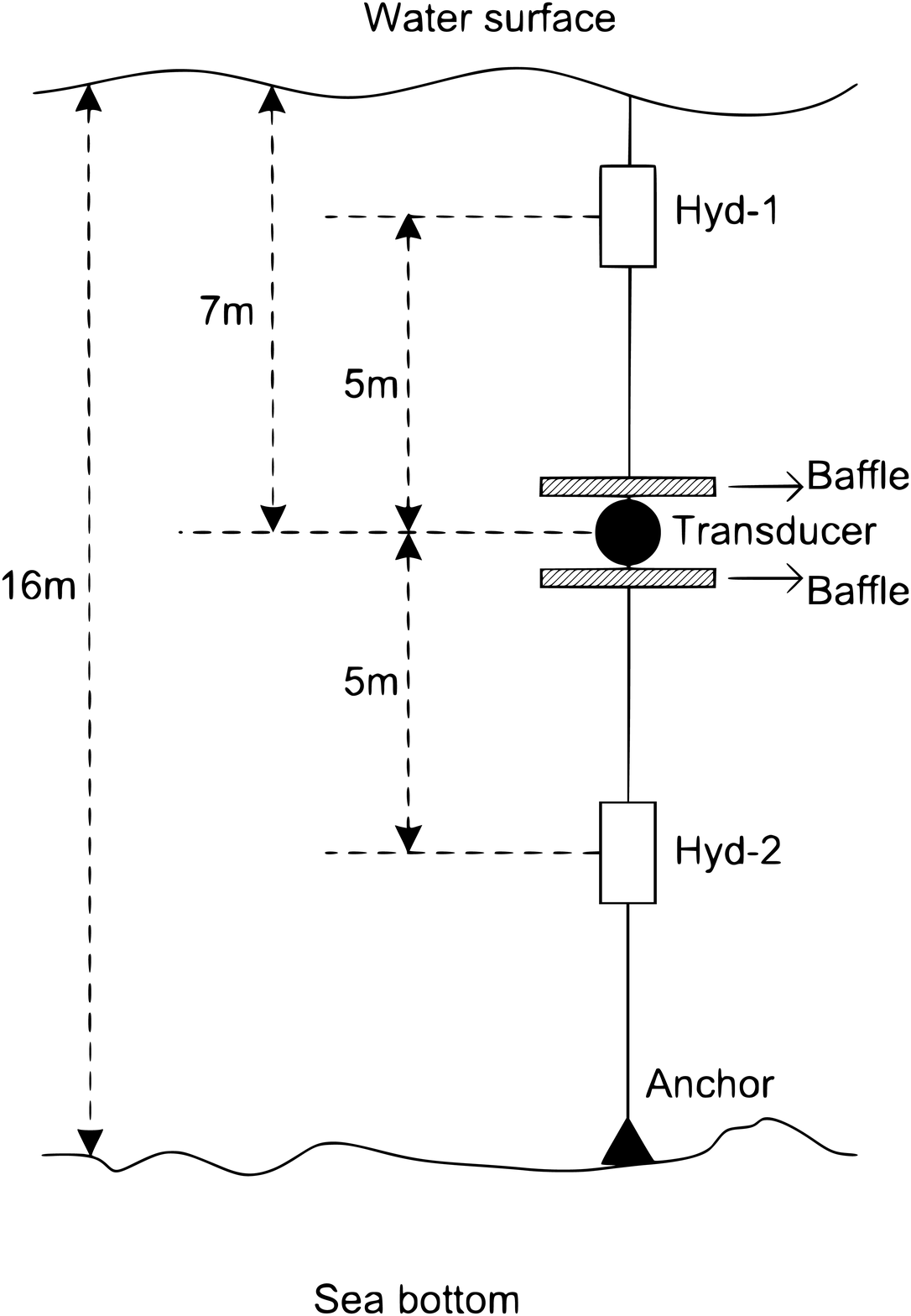}
  \captionof{figure}{Experimental set-up.}
  \label{F2}
\end{minipage}%
\begin{minipage}{.2\textwidth}
  \centering
  \includegraphics[height=2.5in]{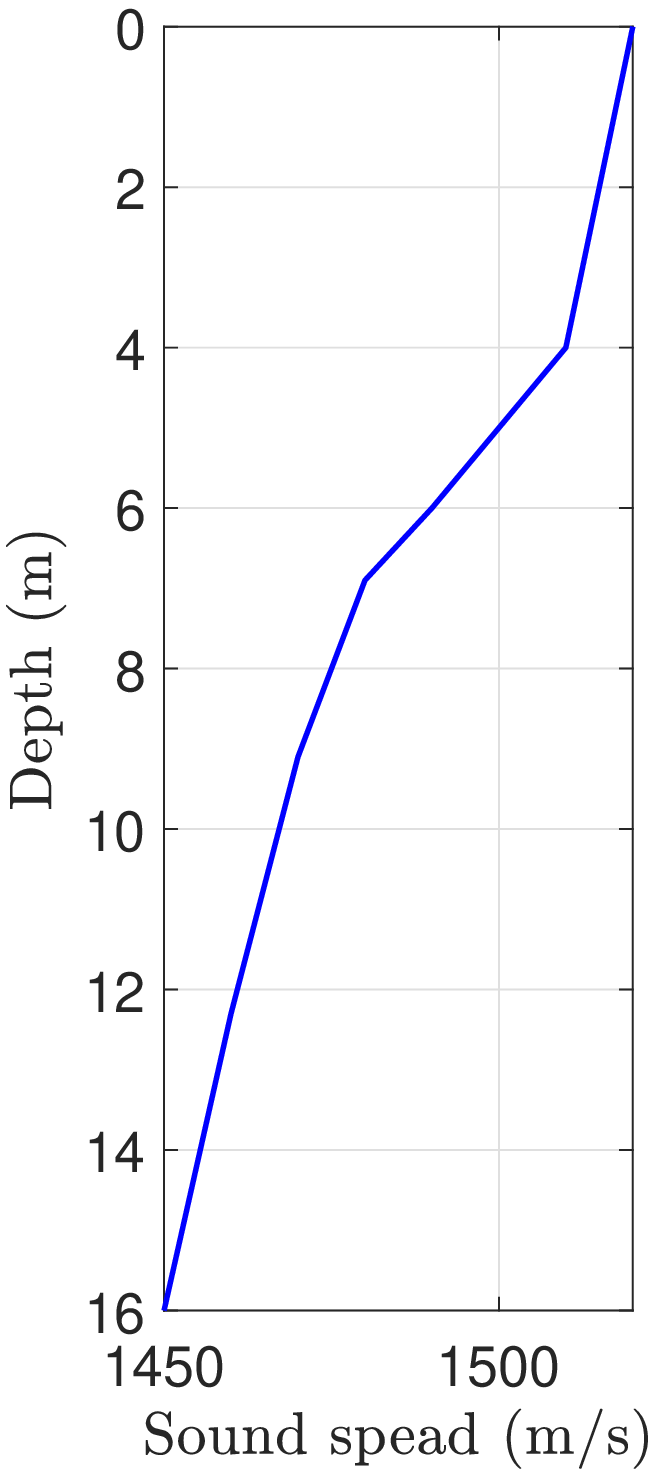}
  \captionof{figure}{Sound speed profile (SSP).}
  \label{F3}
\end{minipage}
\end{figure}

The experiment was performed in Lake Tuscaloosa, AL, USA. Fig.~\ref{F2} illustrates the experimental set-up for one side of the UWA-IBFD communication. In this set-up, the transducer and two hydrophones are  installed linearly on a single pole. The first hydrophone (Hyd-1) is close to the water surface, the second hydrophone (Hyd-2) is close to the bottom and the transducer is located in the middle of hydrophones, $7$ m below the water surface. In order to attenuate the SI, two baffles are installed around the transducer to isolate it from the hydrophones. At the time of the experiment, the wind speed was very low and the water surface was quite calm.
Fig.~\ref{F3} also shows the measured sound speed profile (SSP) of the water at the time of the experiment.

The OFDM signal parameters used in the experiment are given in Table 1. The  transmitted signal contains $N=720$ OFDM blocks with no pulse shaping. This signal takes about $74$ s for transmission. 
The two hydrophones shown in Fig.~\ref{F2} record the received signal with a sampling rate ${f_{\rm{s}}} = 512$ kHz. The inaccuracy of the hydrophones' sampling rate is about $5$ Hz, which is compensated at the receiver. In addition, about $0.3$ Hz of carrier frequency offset (CFO) is removed from the received signals. It should be mentioned that the cut-off frequency of the low pass filter shown in Fig.~\ref{F1} is $5$ kHz.
After pre-processing, the resultant signal is applied to estimate the SCIRs using the FD and TD methods, as in Section II.

\setcounter{table}{1}
\renewcommand{\tablename}{Table} % Set the tablename to Panel, instead of Table
\renewcommand{\thetable}{1} % Setting the table number output to letters 
\begin{table}
\caption{OFDM parameters}
\centering
\begin{tabular}{ |c|c|c|c|c|c|c|c|c|c|c|c|c|c| } 
\hline 
Modulation & BPSK \\ 
\hline
Bandwidth & $B = 5$ kHz  \\ 
\hline
Sampling time & ${f_{\rm{s}}} = 512$ kHz\\ 
\hline
\# of OFDM subcarriers & $K = 265$  \\ 
\hline  
CP length & $\upsilon  = 256$ \\
\hline
OFDM block length & $K + \upsilon  = 512$ \\
\hline
OFDM block duration  & $(K + \upsilon )/B = 102.4$ ms \\
\hline
Carrier frequency  & ${f_{\rm{c}}} = 28$ kHz\\ 
\hline 
\# of OFDM blocks  & $N=720$ \\ 
\hline
\end{tabular}
\label{T1}
\end{table}

\vspace{-5pt}
\subsection{The Estimated PDP}
\begin{figure}[!t]
\centering
\includegraphics [width=3.9in]{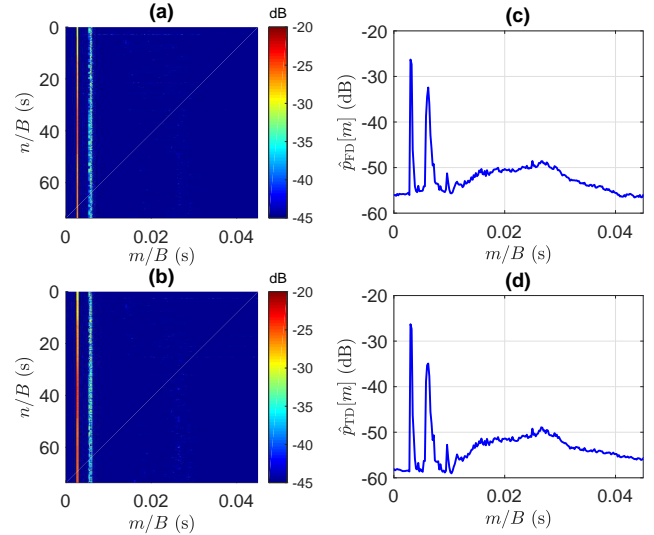}
\vspace{-5mm}
\caption{The estimated SCIR for Hyd-1 using  (a) FD method and (b) TD method. The estimated PDP using  (c) FD method and (d) TD method. In the TD method, the window size is ${L_{{\rm{win}}}/B}=104.2$ ms.} \vspace{-6mm}
\label{F4}
\end{figure}

In Fig.~\ref{F4}, we compare the estimated SCIRs and PDPs for Hyd-1 obtained by using the FD and TD estimators. For the TD method, the observation window size is chosen equal to the OFDM block length (${L_{{\rm{win}}}}={L_{{\rm{blk}}}}=512$) and the channel length is equal to half of the window size (${M} = 256$). 
As can be seen, both methods lead to approximately the same results. The estimated PDP has three components:

1) A principal path at $3$ ms which, according to the experimental set-up and the SSP, is caused by the $5$ m long direct path between the transducer and the hydrophone. 

2) The second largest return at $6$ ms, is the first bounce from the water surface, which travels a total of $9$ m from the transducer to the hydrophone. The surface bounce path is about $9.3$ dB below the direct path.

3) The remaining indirect paths between $14$ ms and $35$ ms are caused by  reflections from the bottom. 

\begin{figure}[!t]
\centering
\includegraphics [width=3.9in]{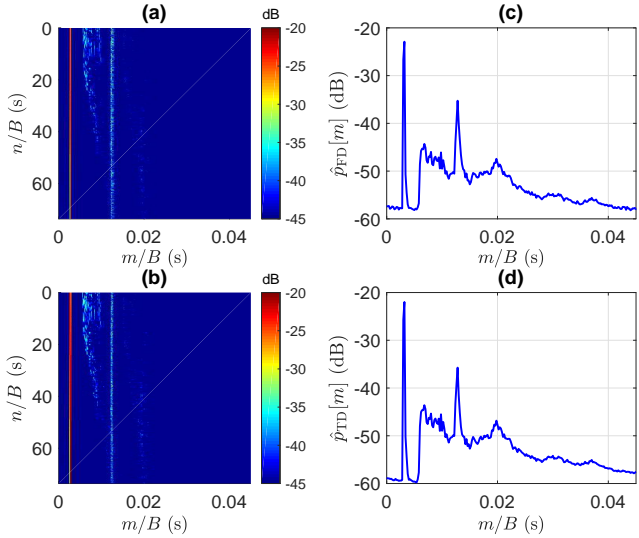}
\vspace{-5mm}
\caption{The estimated SCIR for Hyd-2 using (a) FD method and (b) TD method. The estimated PDP using  (c) FD method and (d) TD method. In the TD method, the window size is ${L_{{\rm{win}}}/B}=104.2$ ms.} \vspace{-5mm}
 \label{F5}
\end{figure}

In Fig.~\ref{F5}, we present the results for Hyd-2. The estimated SCIRs and  PDPs are exactly what we expect based on the position of Hyd-2. It is obvious that the second main return is from the first bounce off the surface, which arrives after the bottom reflections because Hyd-2 is closer to the bottom. This result indicates that, although the first surface bounce has more delay than the bottom reflections, it is still the dominant indirect path of the SCIR.
As a result, one can conclude that, regardless of the hydrophone's depth, the most prevalent indirect path is the first bounce from the surface. 


\vspace{-5pt}
\subsection{The Estimated ACF and COT}

\begin{figure}[!t]
\centering
\includegraphics [width=2.5in]{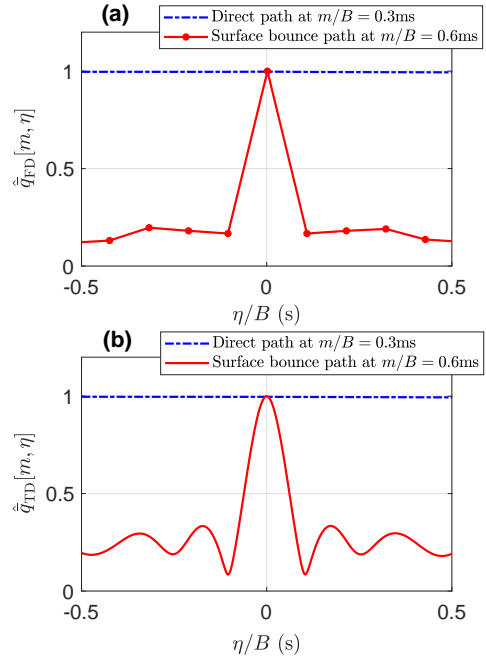}
\vspace{-3mm}
\caption{The estimated normalized ACF of the direct and surface bounce  paths for Hyd-1 using  (a) FD method and (b) TD method. In the TD method, the window size is ${L_{{\rm{win}}}/B}=104.2$ ms.} \vspace{-6mm}
 \label{F6}
\end{figure}

\begin{figure}[!t]
\centering
\includegraphics [width=2.5in]{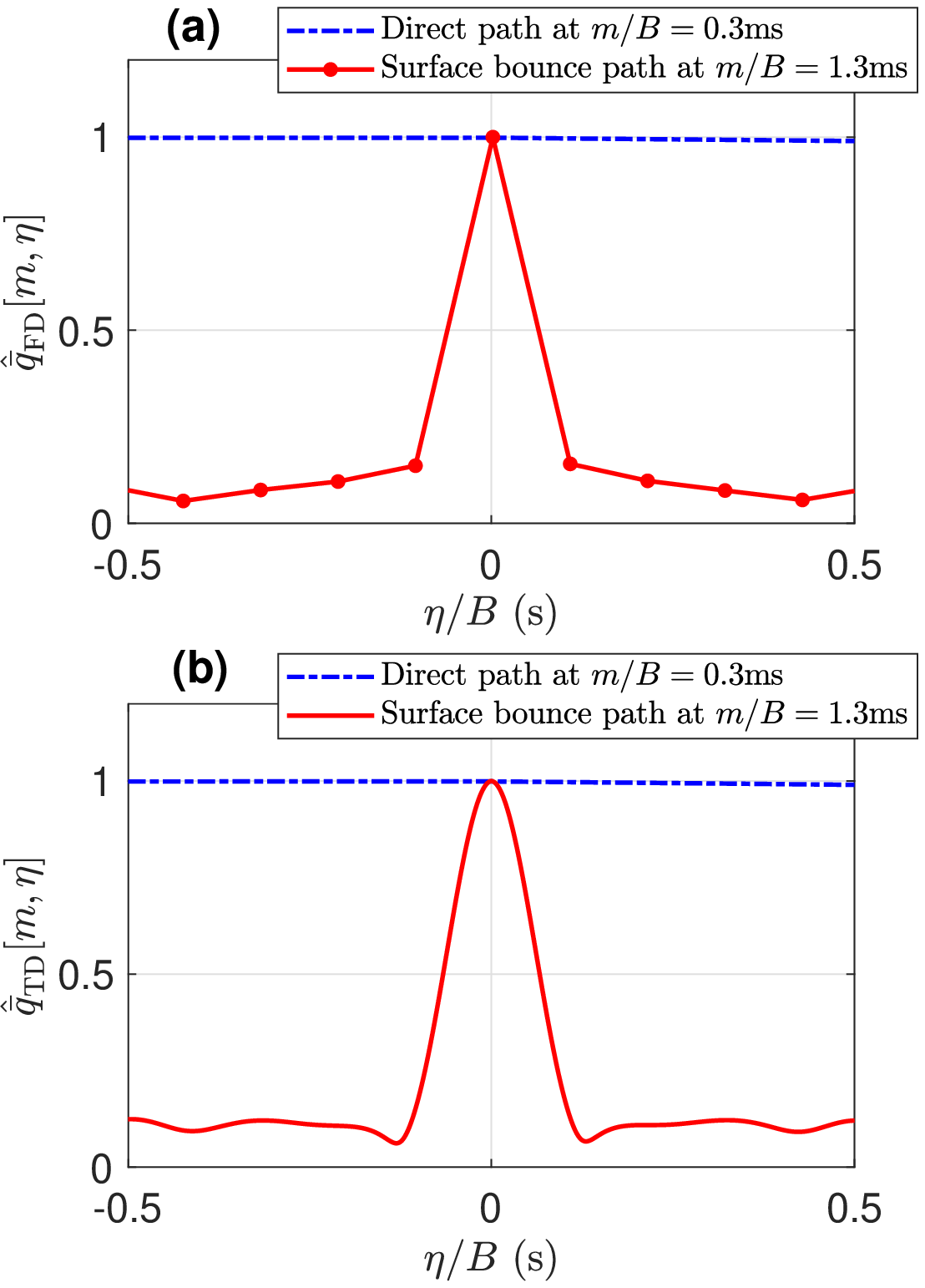}
\vspace{-3mm}
\caption{The estimated normalized ACF of the direct and surface bounce  paths for Hyd-2 using the (a) FD method and (b) TD method. In the TD method, the window size is ${L_{{\rm{win}}}/B}=104.2$ ms.} \vspace{-5mm}
 \label{F8}
\end{figure}

In this section, we use the estimated SCIRs to obtain the ACF and COT of the channel during the experiment. For Hyd-1, Fig.~\ref{F6} indicates the normalized ACF for the two main paths in Fig.~\ref{F4}, including the direct path (at $3$ ms) and the first surface bounce path (at $6$ ms). Notice that the estimated ACFs for each method are identical; however, as discussed in Section II-F, the ACF obtained from the TD method is more accurate (for the FD method, the accuracy is ${L_{{\rm{blk}}}}/B = 104.4$ ms; while, for the TD method it is $1/B=0.2$ ms). Based on these results, the direct path is stable and almost constant over the observation period; on the other hand, the surface bounce path is changing very rapidly. Since the ACF achieved by using the TD method is more accurate, it can be used to calculate the precise COT. Thus, by looking at Fig.~\ref{F6}-b, the COT for the surface bounce is ${\eta _{{\rm{co}}}}/B \approx 72$ ms. 
For Hyd-2, the normalized ACFs are given in Fig.~\ref{F8}. In this case, again the direct path is stable and the surface bounce path is rapidly time-varying, with the COT equal to $\eta _{{\rm{co}}}/B \approx 74$ ms.

\begin{figure}[!t]
\centering
\includegraphics [width=2.5in]{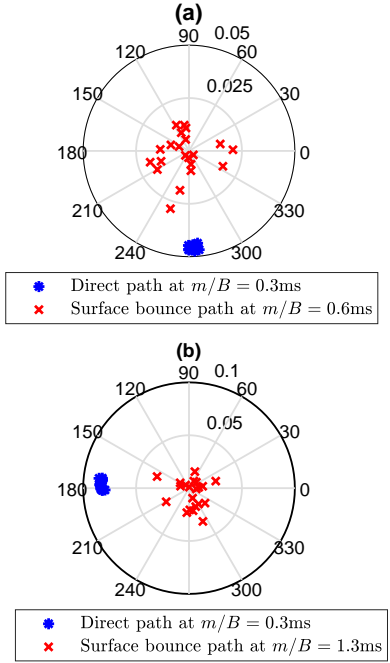}
\vspace{-3mm}
\caption{Polar plot of the direct and surface bounce paths for the SCIR estimated by the FD method (a) Hyd-1 and (b) Hyd-2.} \vspace{-3mm}
 \label{F9}
\end{figure}

In summary, the direct path is relatively stable because the transmitted signal travels a straight and short way to the hydrophone; however, the surface bounce path travels a longer way and is affected by surface fluctuations and the inhomogeneity of the water. 
To obtain more intuition about the rates of change, Fig.~\ref{F9} gives a polar plot of the direct and surface bounce paths in Hyd-1 and Hyd-2, over about $2$ s. Clearly, the direct path changes very slowly; in contrast, the surface bounce is changing very rapidly.

\vspace{-5pt}
\subsection{Discussion}
There are two important points about the results that require some discussion:

1) As mentioned before, the COT of the surface bounce is very short, even shorter than the OFDM block length in the FD method and the window size in the TD method (note that we chose $ {L_{{\rm{win}}}}= {L_{{\rm{blk}}}}$). This means that the assumption of a constant channel over ${L_{{\rm{blk}}}}$ and $ {L_{{\rm{win}}}}$, made in Section II-C and D, is not accurate.  However, recall that we employed these methods just to estimate the statistical characteristics of the SCIR. Thus, because of long-term averaging, this inaccuracy does not affect the estimation of PDP, ACF and COT.
To show that the estimated characteristics are not significantly affected by the aforementioned issue, we shortened the window size in the TD method from ${L_{{\rm{win}}}}=512$ to ${L_{{\rm{win}}}}=180$ (and accordingly, the channel length is taken as ${M} = {L_{{\rm{win}}}}/2 = 90$). Since ${L_{{\rm{win}}}}/B=36$ ms, it is guaranteed that the surface bounce path remains relatively constant over the observation window and the assumption of a steady channel over the observation window is accurate.

\begin{figure}[!t]
\centering
\includegraphics [width=2.5in]{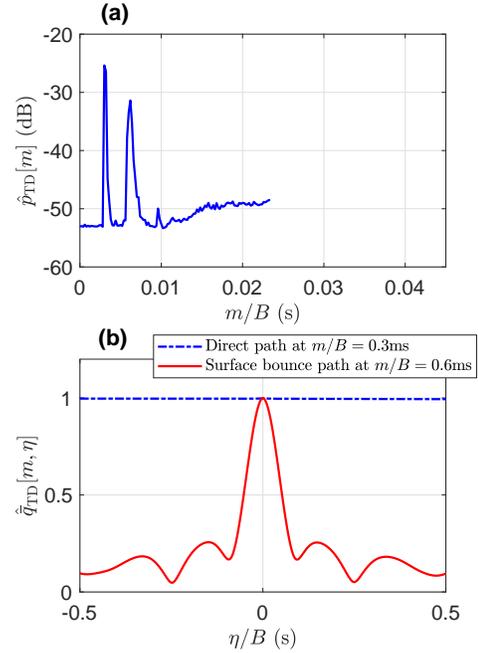}
\vspace{-3mm}
\caption{(a) Estimated PDP and (b) ACF of direct and surface bounce paths using the TD method with window size ${L_{{\rm{win}}}/B}=36$ ms.} \vspace{-3mm}
\label{F7}
\end{figure}

The corresponding PDP and ACF results for Hyd-1 are presented in Fig.~\ref{F7}.
 According to Fig.~\ref{F7}-a, the PDP is exactly what we expect and in  Fig.~\ref{F7}-b the COT of the surface bounce path is similar to what we previously obtained ${\eta _{{\rm{co}}}}/B \approx 72$ ms. The approaches in Fig.~\ref{F7} verify the statement that, although the assumption of a constant channel over the OFDM block length and the window size in Section III-B and C is not accurate, it does not affect the SCIR statistical characterization significantly.   

\begin{figure}[!t]
\centering
\includegraphics [width=2.5in]{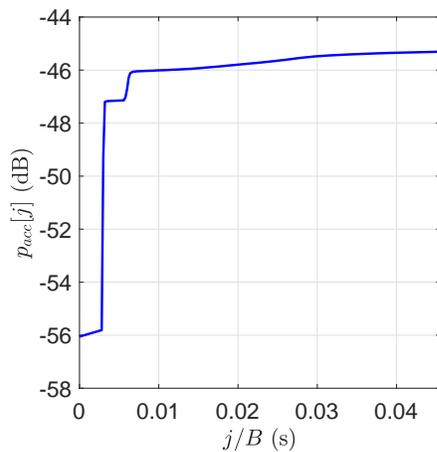}
\caption{Accumulated power of  SI signal for Hyd-1.} \vspace{-4mm}
 \label{F10}
\end{figure}

2) According to the results in this section, the direct path is strong and stable and, as such, easily eliminated. However, the second main path, caused by the surface bounce, is varying rapidly and needs more efforts to track so that one can cancel the SI.
This behavior is also reported for remote UWA channels in half-duplex systems in \cite{R22}; however, it must be noted that  in UWA-IBFD, because the SI is high power (compared to the weak remote signal), variations in the SCIR are more critical.
To investigate how different paths contribute to the SI power, assume that the transmitted signal, $x[n]$, is i.i.d. with power $P_x=1$. According to \eqref{1}, the SI signal power is then given as \cite{R14}
\begin{equation}
\begin{array}{l}
{P_s}{\rm{ = }}{{\rm{E}}}\left\{ {{{\left| {s[n]} \right|}^2}} \right\} = \frac{{{P_x}}}{{{M}}}\sum\limits_{m = 0}^{{M} - 1} {{{\rm{E}}}\left\{ {{{\left| {h[m,n]} \right|}^2}} \right\}} {\mkern 1mu} {\mkern 1mu} {\mkern 1mu} {\mkern 1mu} {\mkern 1mu} {\mkern 1mu} \\
\,\,\,\,\,\, = \frac{1}{{{M}}}\sum\limits_{m = 0}^{{M} - 1} {p[m]}.
\end{array}
 \label{16}
\end{equation}

\noindent where $p[m]$ denotes the PDP of SCIR as defined in Section II. To investigate the contribution of the SCIR delay paths to the total SI power, as in \eqref{16}, we define the accumulated power of  SI as 
\begin{equation}
{p_{{\rm{acc}}}}[j] = \frac{1}{M}\sum\limits_{m = 0}^{j  - 1} {p[m]},
 \label{17}
\end{equation}

\noindent which denotes the power of SI corresponding to the delay paths from $m=0$ to $m=j-1$. In this regard, Fig.~\ref{F10} shows ${p_{{\rm{acc}}}}[j]$ for Hyd-1  using the estimated PDP from Fig.~\ref{F4}.
As can be seen, about $72\%$ of the SI power is from the direct path. In other words, $72\%$ of the SI is easily eliminated. On the other hand, the surface bounce  path includes $16\%$ of the SI power, which is difficult to track. These results indicate that, if the UWA-IBFD system can accurately track just the direct and surface bounce paths, it can eliminate $88\%$ of the SI. For Hyd-2, this percentage is also about $88\%$.
It is noteworthy that because of the rapid changes, the well-designed half-duplex UWA OFDM channel estimators in \cite{R17, R18, R19, R23} may not lead to an accurate SCIR estimation in UWA-IBFD because in these methods the channel is assumed constant over one OFDM block, which we have shown is not correct.

% Can use something like this to put references on a page
% by themselves when using endfloat and the captionsoff option.

\section{Summary}
In this paper, we characterize the self-interference channel impulse response of UWA-IBFD system, using an OFDM signal in both the frequency and time domains. We use the experimental data recorded in a lake water and determine  the power delay profile, the autocorrelation function and the coherence time. According to our results, regardless of the depth of the hydrophones, there are two principal paths in the self-interference channel, including the direct and surface bounce paths. These two paths contain about $88\%$ of the self-interference power. We show that the direct path is powerful and stable and easily eliminated; however, the surface bounce path is weaker and changing rapidly. Having a very short coherence time, the surface bounce path requires more efforts for cancellation.

%\ifCLASSOPTIONcaptionsoff
% \newpage
%\fi

\bibliographystyle{IEEEtranTCOM}

\bibliography{Refrences}

\end{document}